\documentclass[]{iopart}

\usepackage{iopams}  
\usepackage{epsfig}

\def\Journal#1#2#3#4{{#1} {\bf #2}, #3 (#4)}

\def\NPA{{\em Nucl.~Phys.} A}
\def\PLB{{\em Phys.~Lett.}  B}
\def\PRL{\em Phys.~Rev.~Lett.~}
\def\PRD{{\em Phys.~Rev.} D}
\def\PRC{{\em Phys.~Rev.} C}
\def\ZPC{{\em Z.~Phys.} A}
\def\ZPC{{\em Z.~Phys.} C}

\def\EPJC{{\em Eur.~Phys.~J.} C}
\def\PRep{\em Phys.~Rep.~}

\def\JP{{\em J.~Phys.} G}

\newcommand{\expl}{\langle \!\langle}
\newcommand{\expr}{\rangle \!\rangle}

\begin{document}

\title[]{Strangeness in strongly interacting matter}

\author{Carsten Greiner$^\dag$
\footnote{carsten.greiner@theo.physik.uni-giessen.de}
}

\address{$^\dag$ Institut f\"ur Theoretische Physik I, 
               Heinrich-Buff-Ring 16, D-35392 Gie\ss en}

\begin{abstract}
This talk is devoted to review the field of
strangeness production in (ultra-)relativistic heavy ion collisions
within our present theoretical understanding.
Historically there have been (at least) three major ideas for the interest in
the production of strange hadronic particles: (1) mass modification
of the kaons in a (baryon-)dense environment; (2) (early) $K^{+}$-production
probes the nuclear equation of state (EoS); (3) enhanced strangeness production
especially in the (multi-)strange (anti-)baryon channels
as a signal of quark gluon plasma (QGP) formation.
As a guideline for the discussion
I employ the extensive experience with microscopic hadronic transport
models.
In addition, I elaborate on the recent idea of antihyperon
production solely by means of multi-mesonic fusion-type reactions.
\end{abstract}




%
\section{Introduction: The original {\em strange} ideas
or `how we got where we are'}
\label{sec:Intro}

I want to start with a citation of the introduction of an over twenty year
old paper by Randrup and Ko \cite{RK80}: `Since the threshold for their
production is relatively high on the scale of presently availabe beam
energies, the kaons are predominantly produced before the initial motion
is substantially degraded. They are therfore expected to be better suited
as messengers of the primary violent stage of the collision which might
otherwise remain quite elusive.'

This statement is robust and I will come back to it throughout this review.
Although not intended at that former time,
today even up to CERN-SPS energies our strong feeling is
that the overall strangeness
in heavy ion collisions is being produced at the beginning of the
reaction when the system is still far from being near to any
(quasi-)equilibrium stage.

A little later, in the mid-eighties, when the Bevalac had
succesfully started since a couple of years with their relativistic
heavy ion physics program and the AGS as well as the SPS era were on its way,
strange particles had been proposed as an intriguing diagnostic probe
for a variety of interesting questions and phenomena. In the
following I want to list three of them which I personally consider
as the potential `smoking guns' for the then evolving field of
`strangeness in strongly interacting matter' probed
with relativistic heavy ion physics.

\begin{figure}[htb]
\centering\epsfig{file=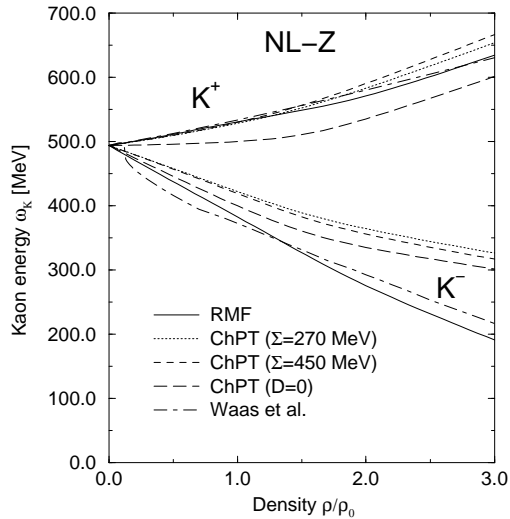,height=8cm}
\caption{The energy of kaons and antikaons situated at rest in
cold nuclear matter obtained within a relativistic
mean field calculation (taken from Schaffner et al \protect\cite{JS97}).}
\label{fig:JS}
\end{figure}

{\em Strange Goings in Dense Nuclear Matter} \cite{KN86}: It was raised
by Kaplan and Nelson that the kaons do feel strongly attractive
(scalar) potentials in the background of a dense nuclear environment
due to the KN-sigma term, resulting in a significant
lowering of their mass with increasing baryon density.
As higher baryonic densities do occur either in
relativistic heavy ion collisions or in the deep interiour of neutron stars,
it was furthermore proposed that
such an attraction actually can ultimately lead to the condensation of
kaons \cite{KN86}.
In addition, there exists also a vector-type interaction given in leading
order (within an effective chiral perturbative expansion) by the
Tomozawa-Weinberg term, which acts repulsive for the kaons and further
attractive for the antikaons. Hence, one was led to the conclusion that
the antikaons should feel a stronger attraction whereas the kaons might feel
a slight repulsion, if at all, with increasing nuclear density. Such a
behaviour, predicted by a mean field calculation, is depicted in
fig. \ref{fig:JS}. Here the free parameters were fixed to the
KN scattering length \cite{JS97}. However, one should keep in mind that
because of the large kaon mass and momenta involved
within a chiral expansion higher order terms
might become significant already at moderate
nuclear densities.
Be it as it is, to leading order such an effect for
the antikaon is a nice demonstration that hadronic particles might significantly
change their property if situated in a nuclear environment.
Modifications of the properties of hadronic particels are eagerly
looked after experimentally in a variety of different
nuclear experiments ($ \gamma , \, e^-,\, \pi ,\, p \, +\, A$ setups
and heavy ion collisions)
as they would prevail new knowledge about the
manifestation of strong interactions in the nuclear medium. In heavy ion
collisions a lowering of the threshold of kaon pair production compared
to the vacuum via the reaction
\begin{equation}
\label{kpair}
N+N \, \rightarrow  \, N+N+ K^+ + K^-
\end{equation}
should lead to a dramatically enhanced production of antikaons especially
for subthreshold energies. This indeed has been observed experimentally.
Is this a clear hint for strong attractive
potentials for the antikaons?

{\em Subthreshold Kaon Production as a Probe of the Nuclear Equation of State}
\cite{AK85}:
It has been shown by Aichelin and Ko within a premature microscopic
Boltzmann-Uehling-Uhlenbeck transport approach that the number of
produced kaons can vary by a factor of $\sim 3$
at subthreshold energies for central collisions
(subthreshold with respect to $N+N\rightarrow N+K+\Lambda $),
depending on the stiffness of equation of state
employed. To understand the idea, one first has to accept that most of the
$K^{+}$s are produced via a second step process
by the more massive $\Delta $s via the reaction
$\Delta + N \rightarrow N+K+\Lambda $ as here the threshold is strongly
diminished. The $\Delta $s in turn have been produced initially via
inelastic nucleon-nucleon scattering. The argument now goes as follows:
For a softer EoS {\em larger} nuclear densities are being buildt up,
{\em lowering} the mean free path $\lambda_{\Delta }$, thus leading to more
inelastic collisions of the $\Delta $, e.g.
\begin{equation}
\label{DelK}
\Delta + N  \,  \rightarrow \,  N+K+\Lambda \, ,
\end{equation}
and hence to more kaons.
This proposal has lead to a considerable interest in the physics of kaon
production at subtreshold energies, in particular, because of the chances
to survey the nuclear EoS being the sought after `holy grail' at that days.
Considerable progress in the concepts of microscopic transport approaches
and in the various reaction channels had been achieved so far. In addition
the kaons might feel a slight repulsion. What
is the status of this idea today?

\begin{figure}[htb]
\centering\epsfig{file=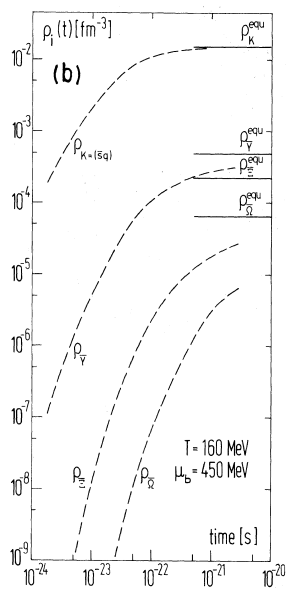,height=8cm}
\caption{Chemical equilibration of antihyperons
and kaons as a function of time in a thermal hadronic gas
taken from the seminal report by Koch, M\"uller and Rafelski
\protect\cite{KMR86}.}
\label{fig:KMR}
\end{figure}

{\em Strangeness Production in the Quark Gluon Plasma} \cite{MR82}:
As the first example, strangeness enhancement
has been predicted  a long time ago by M\"uller and Rafelski as a
diagnostic probe to prove for the short-time existence of a QGP.
The main quantitative idea is that the strange (and antistrange) quarks are
thought to be produced more easily via the fusion of gluons
and hence also more abundantly
in such a deconfined state as compared to the production via
highly threshold suppressed inelastic hadronic collisions.
Enhanced strangeness production via the QGP should thus
been seen most easily in the yield for the most dominant strange particles,
the kaons.
Here, it was argued, however,
that a factor of 2-3 enhancement in the $K/\pi $-ratio relative to the one
obtained in p+p collisons
can only be seen as an indirect signal for QGP creation \cite{KM86}.
With respect to the high production thresholds in the
various binary hadronic reaction channels,
especially the antihyperons and also the
multistrange baryons were then advocated as the
appropriate candidates \cite{KMR86}.
Fig.~\ref{fig:KMR} intriguingly
shows the approach to chemical equilibrium
as a function of time for the various
particle densities of strange hadrons
containing at least one antistrange quark within
a hot and baryonrich hadronic system. Even after 1000 fm/c the antihyperons
do not approach by far their chemical equilibrium values.
It was argued that for a thermalized fireball of hadronic particles
the strange antibaryons are mainly produced via
subsequent binary strangeness exchange reactions with the
kaons, e.g.
\begin{equation}
\label{exchange}
K + \bar{p} \,  \leftrightarrow \,  \pi + \bar{\Lambda } \, ,
\end{equation}
with low crossections.
On the other hand, assuming the
existence of a temporarily present phase of QGP and
following simple, statistical coalescence estimates
the abundant (anti-)strange quarks
can easily be redistributed to combine with the light (anti-)quarks to form the
strange (anti-)baryons \cite{KMR86}, which do then,
in return, come close to their chemical equilibrium values.
Such a behaviour of nearly chemically saturated populations of the
(multi-)strange (anti-)baryons has been experimentally demonstrated
with the Pb+Pb experiments at CERN-SPS.
(This statement can, of course, only be made by invoking an analysis
by a thermal model and fitting the thermal parameters
to the set of individual hadronic abundancies \cite{BMS96,CR00}.
Indeed, the extracted
value for the temperature parameter is significantly close to the critical
temperature $T_C$ obtained by thermodynamical QCD lattice calculations.)
However, Leupold and myself have recently conjectured that
a sufficiently fast and simple redistributions of strange and light quarks
into (strange) baryon-antibaryon pairs might equally well be achieved
by multi-mesonic fusion-type reactions of the type
\begin{equation}
\label{mesfuse}
n_1\pi + n_2 K \, \leftrightarrow \, \bar{Y}+p
\end{equation}
for a moderately dense hadronic system \cite{GL00}.
The beauty of this argument lies in the fact that
(at least) these special kind of multi-hadronic
reactions have to be present because of the fundamental principle of detailed
balance. Are (anti-)hyperons thus still a good diagnostic
tool for the onset of QGP formation in central ultrarelativistic heavy
ion collisions?

In the following sections I will continue
to elaborate in more theoretical detail on the
understanding of strangeness production in relativistic
heavy ion collisions and the three questions raised.
Strangeness production can best be described microscopically
within a hadronic transport approach. This is my personal prejudice.
Therefore, as a basic guideline for the discussion,
I will employ our extensive experience with such a description.
By detailed comparison to data these approaches should provide theoretical
evidence whether each of the above ideas does manifest in nature or not.
As the basic input are vacuum cross sections
for the various binary hadronic reaction channels,
transport approaches do have
a profound and solid foundation: {\em relativistic kinetic theory }
(being, in principle, superior to any thermal or hydrodynamical description)
and known vacuum physics.
However, caution comes into the description
and the character of a model
enters when cross sections are asked which can not be measured experimentally.
Besides the dynamics of all kinds of hadronic resonances this is, in particular,
true for the modelling of the highly energetic inelastic binary hadronic
reactions (by e.g. string fragmentation) and their collective
incorporation into space-time dynamics.

An alternative and (to a certain part) complimentary theoretical description
of the final hadronic yield
being explored in tremendous detail over the last
years is given by various thermal model analyses as already mentioned above.
Why complimentary? As it cannot make any predictions for the initial phase
and how the (pre-)final hadronic system might have reached
some approximate thermodynamical equilibrium,
it exactly does not rest on any (microscopic) model assumptions.
This description helps to extract global properties and systematics
from experimental data
like, being specific, strangeness saturation in the hadronic abundancies.
Here I want to refer to the recent
review by Redlich \cite{KR01}.

In the next section \ref{sec:SIS}
I discuss the production of kaons at
SIS energies. Section \ref{sec:SPS} then deals with kaon and (anti-)hyperon
production at SPS energies.
I will end with a brief conclusion what I do believe what
has been learned at present.

Although important, I will not touch on p+A physics
(for a recent review containing a brief summary please see \cite{N01}).
I will also not discuss the physics of {\em strangelets} \cite{GKS87} or
{\em multihypernuclear objects} \cite{SGS92}, although these exotic states
would clearly represent `strongly interacting strange matter', if
they do exist in nature. For a review and the experimental situation
I refer to \cite{GS96}. In addition, microscopic studies of
single hypernuclei (please see \cite{KEIL} for some new developments)
and the role of strangeness
on the strucure of neutron stars (see \cite{SCHAFFNER})
deserves attention, but will not be
reviewed here. Last but not least on this list of topics not covered
are the encouraging and brand new results on production of strange hadronic
particles within the RHIC program, which now do ask for
more detailed theoretical studies.

\section{Strange Physics at SIS}
\label{sec:SIS}
The production of kaons and antikaons, respectively, in relativistic
heavy ion collisions
below and around the individual threshold energies up to 2 AGeV
has been adressed with significant progress
over the last ten years by the FOPI and the KAOS
collaboration with the SIS at GSI. It has become clear by now that
the data has to be understood in complete and careful detail within
the various transport models in order to pin down possible in-medium effects
for the antikaons or to extract the nuclear EoS. The main
competing approaches are the QMD model (Nantes), the RQMD model (T\"ubingen),
the URQMD model (Frankfurt), the RBUU or HSD model (Giessen) and the
RBUU model (Stonybrook,Texas). I will concentrate in the following to discuss
the recent progress in theoretical understanding being made, but also point
out some present conflicts, at least to my understanding, which has to be
adressed in order to come to solid conclusions.

As will become evident below one first has to understand the production
mechanism(s) of the kaons before considering the antikaons, being more
rarer and thus more exotic. A detailed analysis in comparison to data
for $K^+$ production was first made by Bratkovskaya et al. \cite{BCM97}.
It was found that the inclusive momentum spectra
for various systems and different bombarding energies can be best described
without employing slightly repulsive in-medium potentials for the kaons.
Inclusion of the potentials would lower the total yield by roughly 30\%.
It should be remarked, though, that the kaon flow on the other hand
seems to be more consistent with data by including a slight repulsion
\cite{BCM97}.
In addition, it was found that the reaction
\begin{equation}
\label{piK}
\pi+ N  \,  \rightarrow \,  K^+ + \Lambda \,
\end{equation}
dominates considerably compared to the $\Delta $-induced production via
(\ref{DelK}).  On the other hand, rather opposite findings have been obtained
recently by Hartnack and Aichelin \cite{HA01}. Here the later reaction
(\ref{DelK}) is the most important for producing the kaons, although
different channels do contribute differently for different energies
and for different masses of the system. They do obtain an agreement
to the rapidity spectra for Ni+Ni at 1.93 AGeV when including a repulsive
potential. It might well be that the main difference between the two approaches
stems from a different parametrization of the, in principle, unknown
production cross section (\ref{DelK}). A comparison between both
implemantations of this channel seems to reveal that at least for the
higher energies (at 1.93 AGeV, where the comparison has been made) this
could be the sole reason:
Implementing the {\em same} parametrization of the cross sections
gives rather identcal results for the kaon rapidity spectra
within the two diffenent transport algorithms \cite{HA01}.
A consistent check also for lower energies and different
systems is, however, {\em mandatory}. In any case, as the reaction
(\ref{DelK}) is of definite relevance, a final consensus
on its parametrizaion should be striven for. This seems to be one of the
major ambiguities at present.

Experimental \cite{STURM} as well as theoretical \cite{FUCHS} efforts
concerning the idea on extracting the nuclear EoS from the
$K^+$ excitation function has been made just recently. Because
of the uncertainty in the one cross section as explained,
it was found that the ratios of $K^+$ spectra
for heavy an light ion reactions is a more robust
indicator \cite{FUCHS,HA01}. Also the
kaon yield for light systems should be less sensitive to
the EoS than for heavy systems.
From the theoretical analysis it turns out that the new data do favor
a {\em soft} (and {\em momentum dependent}) nuclear EoS.
Still, if in addition one
does employ a moderately repulsive in-medium potential for the kaons,
the difference in the yield for the kaons between a soft and a hard EoS
shrinkens from a factor of 3 \cite{AK85} down to only 40\% \cite{HA01}.
Refering to the reasoning in the introduction, this counter effect is
understood as for a soft EoS one produces the kaons at higher
baryonic densities where the repulsive potential then increases the threshold.
Hence, an identification of the EoS
has become more delicate compared to the original
idea. Establishing cross correlations with other observables being
also potentially sensitive to the EoS is definitely needed for.
It is worth mentioning that looking at the standard signal of in-plane and
out-of-plane proton and neutron flow experimental data seems
to be mostly consistent
with a soft and strongly momentum dependent EoS (with emphasis more on
the strong momentum dependence of the potentials) \cite{LCGM00}.
Encouragingly, within this same RBUU code the yields of the kaons
are found to be most consistent by employing this same EoS \cite{Larry}.

\begin{figure}[htb]
\vspace*{5mm}
\centering\epsfig{file=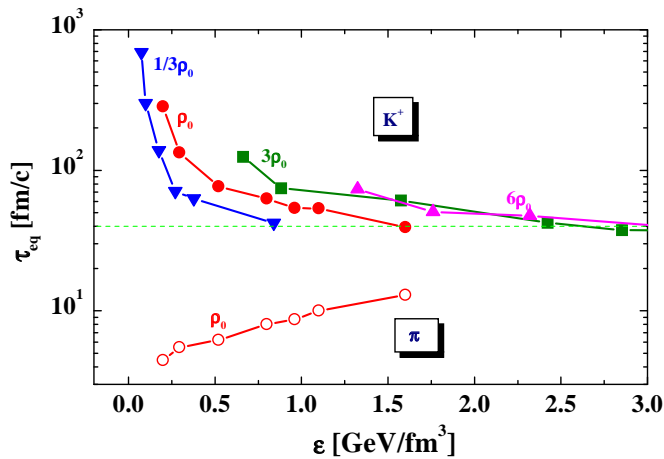,height=6cm}
\caption{Chemical equilibration time versus energy density
for $\pi $ and $K^+$ mesons at various different baryon densities
\protect\cite{BCGEMS00}.}
\label{fig:EB}
\end{figure}

The (subthreshold) production of kaons either via the two step process
of energetic $\Delta $s or pions does happen still at the onset of the reaction
when the buildt up baryon density is high. Afterwards, more or less
no further strangeness is being produced in any of the
microscopic simulations. This reflects the
early statement of Randrup and Ko remarked in the introduction.
Putting it differently, when the
momenta of the nucleons sufficiently degraded and the system
has to some extent thermalized, the timescale for production of
strange particles via the considered kinetic reactions becomes incredibly
large, exceeding two order of magnitude the lifetime of the system.
Fig.~\ref{fig:EB} shows this statement in a dramatic way: Within the
RBUU model the chemical saturation of the kaons has been investigated
microcanonically for a static box with
isosysmmetric baryon content \cite{BCGEMS00}. It is
found that for low energy densities as expected at SIS energies
the saturation time scale exceeds $10^3$ fm/c. (In the figure actually
only energy densities as expected for situation for lower AGS energies
($\sim 4$ AGeV) up to SPS energies are given. For even lower densities
to be expected at SIS see \cite{BCGEMS00}.) A similar conclusion
with a dynamical simulation has been recently shown by Pal, Ko and Lin
\cite{PKL01}. Refering e.g.to the reaction (\ref{piK}),
the fundamental and simple kinetic reason for the large saturation timescale
is the fact that the probability
for a $K^+$ to meet with a $\Lambda $ and then to annihilate becomes
tinely small. It are these kind of annihilation rates of the kaons which
in sum equals the rate towards chemical equilibrium.
Kaon production at SIS energies is an initial nonequilibrium
phenomena as is the overall strangeness production.

It might accidentally be that the amount of produced kaons might ressemble
a chemical equilbrium population at some very late stage of the
reaction as advocated in \cite{COR}.
(The overall population at equilibrium is very sensitive on the temperature
parameter and thus there is a good chance to `fit' to the data with
a particular value. In \cite{PKL01} it was found within the employed
microscopic calculation that at the very late stage there exists temporarily
a situation where the Gain and Loss contribution, although tinely small,
do become equal in size. Although the number of kaons does not change
anymore throughout this final evolution, this particular situation resembles
qualitatively that of a system at equilibrium.)

Being prepared with a good understanding of the production
of the kaons, I want to go over to our present knowledge of
the production of the more exotic probe, the antikaons.
One should be aware that at all SIS energies the kaons and hyperons
are nearly equal in number and the antikaons are much rarer than the
kaons. In principle, there is a good chance that the antikaons
are made by redistributing the strange quarks of the hyperons to
mesonic degrees of freedom. Indeed, it was pointed out nearly 20 years
ago by Ko that the kinetic reaction
\begin{equation}
\label{akpi}
\pi  + Y \, \leftrightarrow  \, N + K^-
\end{equation}
should play a dominant role in the production of subthreshold
antikaons. The cross section of the back reaction is known experimentally
quite well and becomes very large close to threshold.
The direct reaction can then be obtained by means of detailed
balance.
As (\ref{DelK}) is thought to be rather dominant for kaons, one
might think similarly also of a second step process involving the
$\Delta $,
\begin{equation}
\label{akdel}
\Delta   + Y \, \rightarrow  N + N + K^- \, \, ,
\end{equation}
for the production of the kaons \cite{HOA01}.
(Here, however, again the cross section
is not known experimentally and thus has to be modelled.)
Also, as shown in \cite{CBMTS97}, the more direct pion induced reaction
with energetic, incoming nucleons
\begin{equation}
\label{akpiN}
\pi   + N \, \rightarrow  N + K^+ + K^- \, \, ,
\end{equation}
does contribute considerably to the yield, especially when turning on the
attractive potentials \cite{Elena}.

Comparing the two microscopic approaches,
\cite{CBMTS97} and \cite{HOA01}, both dealing with a detailed description of
antikaon production, one has to clearly admit, that a comprehensive
emerging picture is at present still not really given.
Both approaches do have in common, that they show, that the
elementary NN-reaction (\ref{kpair}) in the vacuum is of
{\em no } importance. The antikaons are produced via second or higher step
processes. Within the
RBUU model an attractive potential for the antikaons is definetely
needed in order to describe the various $K^-$-spectra \cite{CBMTS97}.
(The $K^+$ are calculated without any - in this case repulsive - potentials
as this is shown to describe the various data as outlined above.)
Reabsorption on the nucleons, i.e. the backward reaction of (\ref{akpi})
is shown to be important so that it is argued that
a sufficient fraction of kaons has to be
produced via (\ref{akpiN}) when the baryon density is still high
and the mass reduction of the antikaons is at work.
Within the QMD model, however, it is shown
that the reabsorption {\em and} production
via (\ref{akpi}) is by far the most important channel \cite{HOA01}.
Production and Absorption do happen on a very fast timescale and do
occur hand in hand. The mass law action seems to be at work so that
the antikaons and the hyperons are in chemical equilibrium
{\em relative} to each other. (This picture
gives a microscopic justification for the use
of a thermal and chemical equilibrium model \cite{COR} at SIS energies
at least for the antikaons.)
When all particles do freeze out, the
baryon density is rather low, so that the possible attractive potentials
have not any significant influence on the abundance of the antikaons
any longer. Any potential antikaons of the early high-density phase,
when the reduction in mass is of most importance,
will get reabsorbed with a probability near to one
in the ongoing evolution of the system \cite{HOA01}. This view
stands in contrast to the findings and argumentations of \cite{CBMTS97}.

Does one see the medium effects of the antikaons, i.e. a significant
lowering in mass when the baryon density is high?  Yes and no, at present.
The RBUU analysis does require a strong attractive selfenergy for the kaons
to account for the measured yield. QMD does not, although they do
need some repulsive potential for the kaons to get the number in kaons
and thus also in the hyperons correct.
Here, however, as just explained, due to the strong reabsorption
it does not really matter as with or without the potentials
the final yield in antikaons is calculated to be more or less the same
\cite{HOA01}.
The experimental data by KAOS and FOPI have become much refined
in quality over the years and now do ask for a clear theoretical (re-)analysis.
My presumption is that the parametrization and implementation
of the cross sections for the reaction (\ref{akpi}) has to be
accurately cross checked by the various competitors in the field.
It is somewhat suspicious that the one group sees a much stronger
reabsorption of early produced antikaons than the other group.
This might well be only one possible source of some (minor) mismatches.
A certain `{\em unification}' in the various parametrizations
and implementations of cross sections is again {\em mandatory}.

Before finishing the discussion on SIS physics, I want to mention that
indeed the medium effects of the antikaons might not be as simple as
advocated above and as incorporated simplistically by mean field type
potentials in the various transport approaches. It is known for
a long time from coupled channel analyses of $K^-N$-scattering data that the
$\Lambda (1405) $-resonance strongly couples
in the s-channel slightly below threshold to the
two particle system. In medium, roughly spoken, this coupling results in an additional
$\Lambda (1405)$-hole branch for the kaon(-like) excitations with
a nontrivial form for their spectral function \cite{K94,L98} having
a structure much different from a simple quasi-particle.
This is an intriguing topic. Over the years nontrivial possible medium
modifications of spectral functions of e.g. vector-mesons
(because of their electromagnetic decay channel) has attracted
a lot of interest in the heavy ion physics community. However, considering
solely the kaons, it is a delicate question at present whether such effects
really are visible by means of a final yield of onshell antikaons.
In addition a full understanding of a consistent transport scheme
for continous, off-shell particle-like excitations has not yet emerged.

\section{Strange Physics at AGS and SPS}
\label{sec:SPS}
A first and thorough attempt to understand strangeness production
for AGS and SPS energies
within one microscopic model, the RQMD model, was carried
out over many years by Sorge and coworkers. For a nice discussion
of the various results, pre- and postdictions, and physics points of
view I like to refer to \cite{So98}.

\begin{figure}[htb]
\vspace*{1cm}
\centering\epsfig{file=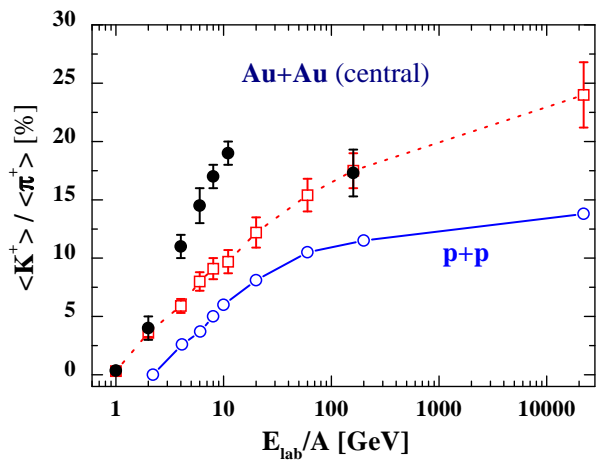,height=6cm}
\caption{Excitation function for $K^+/\pi^+$
around midrapidity
for central Au+Au reactions (open squares) from SIS to RHIC energies
obtained within the HSD model
in comparison to experimental data and elementary p+p collisions (open circles)
\protect\cite{Ge98a}.}
\label{fig:Jochen}
\end{figure}

As outlined in the previous section, certain differences
in the implementations of particular cross sections already had some
profound influence on the production of kaons and antikaons
within the various transport descriptions.
Turning now the discussion to the higher incident bombarding energies,
it should not be a too big surprise that the
predictions for strangeness productions should possibly vary even more.
This is indeed the case, at least for all AGS energies, ranging from
2 to 11 AGeV, and the lower SPS energies, and will be briefly
reported below (see also the contribution of Bass \cite{BASS}).
Dealing with the description of the various inelastic reactions,
a certain physics point of view and hence a true character of
a model enters: The very first nucleonic and hadronic collisions
are highly energetic. Particle production is not directly calculable
and can only be described with phenomenological descriptions being
adjusted to data like the well known Lund model with string fragmentation.
In addition, certain classes of inelastic hadronic reactions at
intermediate energies might still be modelled by various
(baryonic and mesonic) resonances
being not so much well established.

In a recent systematic study the properties of $K^+ $, $K^-$ and
$\Lambda $ particles in nuclear reactions from SIS to CERN-SPS energies
have been investigated
within the microscopic HSD model \cite{Ge98}.
The outcome for the excitation function of the most dominant
strange particles, the $K^+$-mesons,
relative to the $\pi^+$-yield within the HSD model
is summarized in  fig.~\ref{fig:Jochen}.
Since a relative enhancement of strangeness is observed already
in hadron-hadron collisions  for increasing energy
the to be measured strangeness should be compared
relative to p+p collisions at the same energy.
After the primary string fragmentation
of intrinsic p-p--collisions the hadronic fireball starts with a
$K^+/\pi^+$ ratio still far below chemical equilibrium with $\approx 6 - 8 \% $
at AGS to SPS energies before the hadronic rescattering starts.
Secondary
(meson-baryon) and ternary (meson-meson)
induced string-like interactions do then contribute significantly
to additional strange particle production, particular for reactions
at SPS energies.
Via these channels about the same number of strange and anti-strange quarks
is produced as in the primary p+p collisions. This then can explain the
factor 1.75 as the relative enhancement compared to p+p
(compare fig. \ref{fig:Jochen}).
Hence, the major amount
of produced strange particles (kaons, antikaons and $\Lambda $s) at SPS-energies
can be understood in terms of early and still energetic,
non-equilibrium interactions.

As the average kinetic energy and the particle density increases monotonically
with incoming kinetic energy of the projectile while the lifetime of the
fireball increases with the system size, a smooth and continous
enhancement is expected in a hadronic description by these effects.
Experimentally, at AGS energies and also at the lower, new SPS data,
a different bahaviour for the excitaion $K^+/\pi^+$
is seen. The relative enhancement
factor here is $\approx 3$ and can not be fully explained within the cascade
type calculations \cite{Ge98}.
Strangeness production is underestimated by $\approx $ 30\%
compared to explicit kaon data \cite{Ge98} wheras the pion population
is in addition slightly overestimated.

This discrepancy in strangeness production at AGS energies indicates
either hadronic physics not taken into account or
some new nonhadronic physics involved
for the primary $s\bar{s}$ production.
Including kaon potentials does help to some extent to
raise somewhat the production of $K^+$ and especially $K^-$ mesons
at highest AGS energies,
still a significant underestimation at AGS energies
does persist \cite{Ge98a,CB99}.
The above HSD results do agree more or less with older
RQMD results \cite{GH95} (version 1.7)
and ARC results \cite{Ka96}, whereas newer RQMD calcualtions \cite{So98}
(version 2.3) are more in line with data. (Unpublished)
URQMD calculations, as reported
in \cite{BASS}, do find agreement for the lower AGS energies, but
otherwise appreciably do underestimate the ratio.
In the here reported HSD calculations only
established baryon resonances up to
$N^*(1535)$ are included, wheras at least in the newer version
of RQMD \cite{So98} resonances of much higher mass have been implemented.
As for the AGS energies and the lower SPS energies the baryon densities
achieved are the highest, especially then the baryonic resonances
could be of crucial importance and thus
could be the possible reason for the significant
differences in the various numerical simulations.
At the full SPS energies, where the secondary interactions are meson
dominated, the possible influence of baryonic resonances is diminished
and this then could explain why at this much higher energy the various
model predictions do not differ as much as compared to the
various AGS energies.
(The changing role and importance of either baryonic or mesonic
contributions to the $K^+/ \pi^+$ excitation function
when going from AGS to SPS energies has recently
also been obeserved within a thermal model analysis \cite{BCOR02}.)
Clearly, a unified description would be
desirable, however, different physical interpretations for the various
model inputs can hardly be justified by a theoretical
ab initio derivation at present.

Closing the discussion on the outcome of strangeness production
at AGS and SPS energies within transport simulations, I do mention once more
that the overall strangeness is produced by the very first highly
non-equilibrium collisions.
The amount of strange quarks in the system is then roughly conserved
when the energetic and non-equilibrium reactions have ceased and
the system has started to equilibrate locally in its
particle momentum distributions.
This observation is indeed the continuation to much
higher bombarding energies of the early statement raised by Randrup and Ko
for Bevalac energies. Refering again to our recent investigation
\cite{BCGEMS00}, summarized in fig.~\ref{fig:EB},
the chemical saturation timescale of kaons in a hadronic transport
description is larger than 40 fm/c
for all equilibrium energy densities up to 2-3 GeV/fm$^3$,
and thus exceeds considerably the lifetime
of the fireball in the center of mass frame.

What do we know about predictions on strangeness production
in a QGP
or on the strangeness content of a fully saturated QGP?
The basic answer is `{\em not much more}' than
in the days when the first pQCD estimate \cite{MR82} was made.
This is a personal, and provocative statement. The old and
simple estimate for the saturation rate
of strangeness was based on the elementary cross section
$ \sigma _{gg\rightarrow s \bar{s}} \approx \frac{\pi \alpha_s^2}{3s}
\ln \frac{s}{m_s^2} $
of two free gluons
fusing to a strange-antistrange
quark pair
folded with thermal distributions for the gluons.
The cross section depends logarithmically on the value of the
strange quark mass. Instead of using the current masse of $m_s\approx 150 $ MeV
one might consider the typical value of $m_s^{NJL} \approx 400-500$ MeV
obtained within a Nambu-Jona-Lasinio approach at and slightly above
chiral restoration temperature \cite{R99}, yielding roughly a factor of 2-3
smaller cross section in the kinemetical region of interest. The
powerful tool of `hard thermal loop' resummed techniques,
in order to deal with the dynamically generated masses in a
thermal field theoretical, infrared safe and consistent framework,
for calculating
strangeness production in the QGP has not been carried out with full success
and a lot of uncertainties do remain \cite{SH96}.  Lattice gauge
calculations so far have not contributed at all to the question
of how much strangeness is inside a saturated QGP or inside
a hadronic system close to the critical temperature.
The strangeness content is again based solely on simple, perturbative
estimates. Furthermore, hadronization is a nonperturbative
phenomena and it is not known whether strangeness can be produced
or not when hadrons are created out of a deconfined QGP state.

\begin{figure}[htb]
\centering\epsfig{file=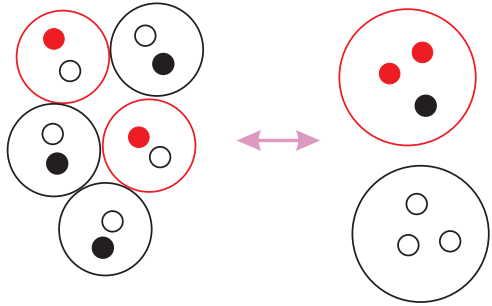,height=4cm}
\caption{Schematic picture for the multi-mesonic fusion-like
reaction $3 \pi  + 2  K \leftrightarrow \bar{\Xi } + N$.}
\label{fig:Fusion}
\end{figure}

As mentioned in the introduction,
a factor of 2-3 enhancement in the $K/\pi $-ratio relative to the one
obtained in p+p collisons was not really considered as an unambiquous
signal for QGP creation.
Strangeness enhancement relative to p+p is seen from SIS energies up to
nowadays RHIC energies. With particular respect to the
experimental findings at SIS and AGS
it is valid to say that strangeness enhancement does
not require deconfinement.
In view of our discussion above,
strangeness production can be understood in terms of hadronic physics,
although this does not disprove a temporary creation of a QGP.

The true proposed and advocated candidates have been the
multistrange baryons, and,
even more strikingly, all antihyperon species \cite{KMR86}.
The analysis of
abundancies of hadronic particles and especially strange particles,
measured by NA49 and WA97,
within thermal models does strikingly support the idea of
having established a (thermodynamically) equilibrated
and chemically saturated hadronic fireball
in some particular late stage of the reaction, dubbed `point'
of {\em chemical freeze-out}
(for analyses of Pb+Pb collisions at CERN-SPS see \cite{BMS96,CR00}).
In this respect especially a nearly fully saturated yield
of antihyperons is found.
In the following I will now
report on our recent idea \cite{GL00}
of rapid antihyperon production by multi-mesonic reactions of the
type (\ref{mesfuse})
(see also the schematic illustration fig.~\ref{fig:Fusion}) and
provide new calculations considerably supporting this
new insight.
This idea does rest on the (conservative) view
that before chemical freeze-out already a
hadronic system has been established.

A few detailed, yet purely phenomenological attempts
to explain a more abundant production of antihyperons within
a hadronic transport description do exist like the color rope formation
by Sorge et al \cite{So95,So98} or the high-dense cluster formation of
Werner and Aichelin within the VENUS transport approach \cite{WA93}.
The underlying mechanisms, however, have to be considered as exotic:
Sorge (as well as other studies like e.g. \cite{BB00}) point out in detail
the dramatic role of antibaryon annihilation, which he more then counterbalances
by initial formation via color ropes. He also finds that
the binary exchange channel (\ref{exchange}) is actually not as small as
advertised originally in \cite{KMR86}, but helps rather quickly to repopulate
the various strange antibaryon populations, if enough antibaryons are present
(compare also the above discussion at SIS energies for the
dominance and importance of the complete similar (T-conjugated)
reaction (\ref{akpi})).
The high-dense cluster formation, on the other hand, resembles
in its philosophy the idea
that in a very dense hadronic system potential multi-particle interactions
could be modelled by a simple statistical treatment.

What are the new arguments?
If the hadronic degrees of freedom are in a state of thermodynamical
equilibrium, which constitutes the basic concept of any thermal
model analyses, a dynamically realization to describe
such a system has to fulfill the concept of detailed balance
in the considered chemical reactions.
As the annihilation of antihyperons on baryons is of dramatic relevance,
the multi-mesonic (fusion-like) `back-reactions'
involving $n_1$ pions and $n_2$ kaons,
corresponding to the inverse of the strong binary
baryon-antihyperon annihilation process (similar to the standard
baryon annihilation $\bar{p} + p \rightarrow n \, \pi $),
have, in principle, to be taken care of in a dynamical simulation.
($n_2$ counts
the number of anti-strange quarks within the antihyperon $\bar{Y}$.
${n}_1+ n_2$ is expected to be around $5-7 $.)
This reasoning has first been raised recently by Rapp and Shuryak
concerning the production of anti-protons \cite{RS00}.

\begin{figure}[htb]
\centering\epsfig{file=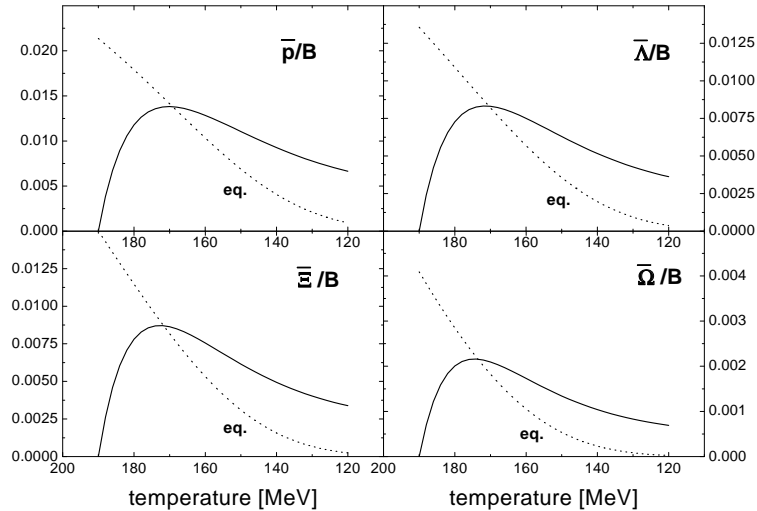,height=8cm}
\caption{The antihyperon to baryon number ratio
$N_{\bar{Y}}/N_B \,  (T)$ and $N_{\bar{Y}}^{eq.}/N_B (T) \,  $
as a function of the decreasing temperature. For the evolving
system an isentropic, Bjorken-like (including transversal)
expansion $V(t)$ has been assumed with an entropy per baryon of $S/A=30$,
yielding the explicit dependence $T(t)$ for the temperature and similarly
for the chemical potentials as functions of time.
$N_{\bar{Y}}(t_0=3\, \mbox{fm/c})$ is set to zero.
}
\label{fig:MasterI}
\end{figure}

The annihilation rate $\Gamma _{\bar{Y}}$, as implemented
in the various codes, is large and it should be
approximately similar in magnitude as
for $p+\bar{p}$
at the same relative momenta. Hence, with
$\sigma _{p \bar{Y}\rightarrow n_1 \pi + n_2 K}
\approx 50 $ mb one has
\begin{equation}
\label{rateinv}
(\Gamma  _{\bar{Y}})^{-1} \,  \equiv  \, 1/
(\expl \sigma _{\bar{Y}N} v _{\bar{Y}N} \expr \rho_N) \,  \approx \, 1-3 \,
fm/c \, ,
\end{equation}
when adopting for the baryon density
a typical value $ \rho_N  \approx 1-2 \rho_0 $
when the system is still more dense than at
the chemical freeze-out `point' with
$ \rho_N  \approx 0.5-1 \rho_0 $ \cite{BMS96,CKW01}.
It is this rapid annihilation process which also
does dictate the timescale
of how fast the antihyperon densities do approach local
chemical equilibrium with the pions, nucleons and kaons.
This follows rigorously from the concepts of kinetic theory
\cite{GL00,CG01}. In principle, one can start from
a kinetic Boltzmann-type equation taken into account these
multi-mesonic fusion-type reactions as the production term
being the necessary counterpart
to the annihilatin term. From the microscopic Boltzmann description
master equations, as presented
in \cite{GL00}, can be obtained in a direct way \cite{CG01}.
Assuming further (for simplicity)
that throughout the later expanison of the fireball
the pions, baryons and kaons do stay in thermal and chemical
equilibrium, a simplified master equation for the number of antihyperons as a function
of time can be written in the most simple, but yet direct
and illustrative form\cite{GL00}
\begin{equation}
\label{masterd}
\frac{d}{dt} \rho _{\bar{Y}} \,  = \,    - \,
\Gamma  _{\bar{Y}}
\left\{
\rho _{\bar{Y}} \, -  \,
\rho ^{eq }_{\bar{Y}}
\right\} \, \, \, ,
\end{equation}
where production due to the multi-mesonic `back-reactions' is hidden in the
second term $\Gamma  _{\bar{Y}} \rho ^{eq }_{\bar{Y}}$.

The annihilation ($\equiv $ chemical saturation) timescale
of the antihyperons is indeed small,
and, according to (\ref{rateinv}), it is roughly proportional to the
inverse of the baryon density.
Still it has to compete
with the expansion timescale of the late hadronic fireball, which
is in the same range or larger. It becomes clear
that these multimesonic, hadronic reactions, contrary to binary
reactions, can explain  most conveniently a sufficiently
fast chemical equilibration of the antihperons.
A redistribution of some strange antiquarks
out of the reservoir of the most abundant strange particles,
the kaons, and of light antiquarks out of the huge reservoir of pions
occurs to populate
the much rarer antihyperon degrees of freedom
to its equilbrium value.
Beyond a certain `point' (which, of course, is actually
some continous regime where inelastic decoupling occurs) with already
a moderately low baryon density (and correspondingly
low pion and kaon densities) it will be that the multi-mesonic
creation process becomes more and more ineffective. This can explain, as
will be demonstrated now, the clear `position' of the chemical freeze-out
for the antihyperons.

\begin{figure}[htb]
\centering\epsfig{file=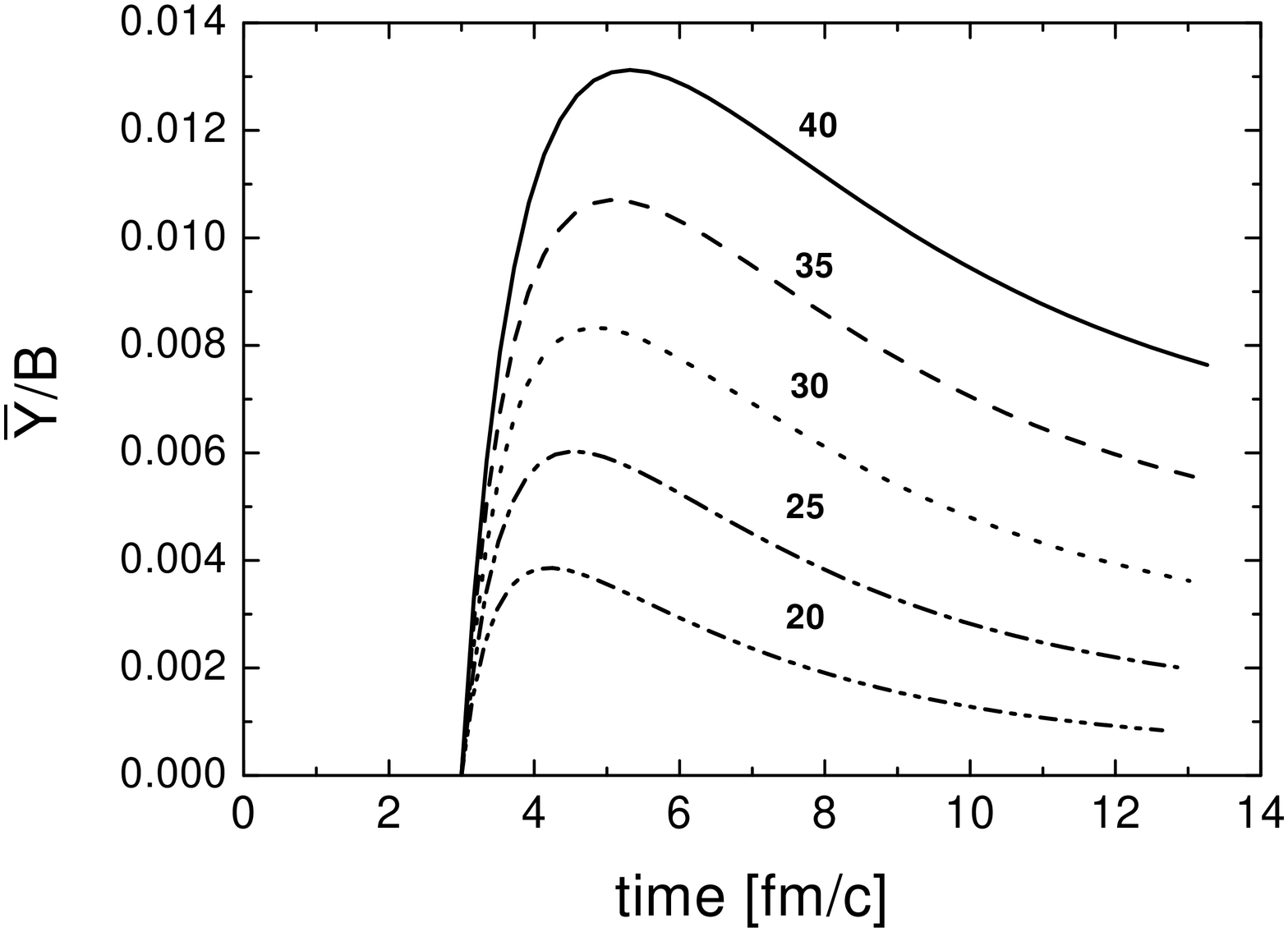,height=7cm}
\caption{The anti-$\Lambda $ to baryon number ratio
$N_{\bar{Y}}/N_B \,(t) $ as a function of time
for an assumed isentropic, Bjorken-like (including transversal)
expansion $V(t)$ and for various entropy content described
via the entropy per baryon ratio ($S/A=20-40$).}
\label{fig:MasterII}
\end{figure}

To be quantitative, two results of a recent study \cite{CG01}
will be briefly explained. The global late-time evoltion of the system
is assumed to be an isentropic expansion of the hadronic resonance gas
with fixed total entropy content being
specified via the entropy per baryon ratio $S/A$.
The effective volume $V(t)$ is parametrized as function
of time by longitudinal
Bjorken expansion and including also an accelerating
radial expansion.
At starting time $t_0 $ an initial temperature $T_0$ is chosen.
(In the following two figures $T_0$ is set to $190$ MeV.
For the various entropy per baryon ratios $S/A =20-40$ this corresponds
to an initial baryon density of $\rho_B (T_0; S/A) = 2.5-1 \rho_0$ and
to an initial energy density of about 1 GeV/fm$^3$.)
From this ansatz the temperature and the chemical potentials do
follow as function of time.
The initial abundancy of antihyperons is set to zero (as a minimal
assumption). A set of master equations of the principal structure as
(\ref{masterd}) are solved.
In fig.~\ref{fig:MasterI}
the number of antihyperons of each specie
(normalized to the conserved net baryon number)
as a function of the decreasing temperature is depicted.
For a direct comparison the instantaneous equilibrium abundancy of antihyperons,
$N_{\bar{Y}}^{eq.}(T(t),\mu_B(t),\mu_s(t))/N_B$, is also shown.
The entropy per baryon is chosen as $S/A=30$ being characteristic
to global SPS results \cite{GKS87,CKW01,Sp98}.
The equilibrium number strongly decreases as a function of
time (or decreasing temperature). The typical
characteristic of the depicted results is that first the antihyperons
are dramatically being poulated, then their individual yield does overshoot
its respective equilibrium number
and then does finally saturate at some slightly smaller value
because of some final reabsorbtion.
From fig. \ref{fig:MasterI} one sees that
the antihyperons effectively do saturate at a number
which can be compared to an equivalent equilibrium number
at a temperature parameter around $T_{eff} \approx 150-160$ MeV.
This value for the temperature is strikingly close
to the ones obtained within the
various thermal analyses \cite{BMS96,CR00,CKW01,Sp98}.
(For the baryochemical potential a similar finding holds \cite{CG01}.)

In fig.~\ref{fig:MasterII} the number of anti-$\Lambda $s
as a function of time is given for various entropy per baryon ratios.
One notices that the final value in the yield significantly
depends on the entropy content, or, in other
words, on the baryochemical potential. Especially the results
at midrapidity from WA97 can best be reprouced by employing
an entropy to baryon ratio $S/A=40$ \cite{CG01}.

Summarizing,
multi-mesonic fusion-type reactions
are a consequence of detailed balance and as
the rate $\Gamma_{\bar{Y}}$ is indeed very large for a
dense hadronic system chemical equilibrium is approached very
quickly.
This is a remarkable observation
as it clearly demonstrates
the importance of hadronic multi-particle channels.
In this respect I would like to mention the recent proposal
by Karliner and Ellis, that an intermediate multi-pion-state
can actually explain the so called $e^+e^-\rightarrow \bar{N} N$ puzzle
at threshold energies \cite{Ka01}.

As the basic assumptions, i.e.
(I) the annihilation cross section
for antihyperons colliding with a nucleon
is roughly as large as the measured one for $\bar{p} + p$ and
(II) at the onset for the equilibration of the antihyperons
one has to assume a somehow equilibrated hadronic fireball
with still a moderate baryonic density,
are quite modest and not of any exotics,
the enhancement of antihyperon production is not a true
QGP signature.
One can always argue that (1)
the hadronic gas can only exist as a dilute sytem
with an energy density to be considerable below 1 GeV/fm$^3$,
otherwise the system would be in a deconfined QGP state,
and/or (2) the expansion goes dramatically rapid
so that any inelastic interactions in the late hadronic system
are non-effective \cite{RL00,Sp98}. But these are just other model
assumptions.

One could be tempted to ask whether a similar reasoning also applies
for the multi-strange hyperons (the $\Xi $ and the $\Omega $) for
which also experimentally a significant enhancement has been reported.
The answer is `no'. The equilibration rate here would be governed by the
density of antibaryons and is thus too low, or putting it differently,
the equilibrium density of multi-strange hyperons is much higher
than the one of antihyperons. Especially the $\Omega $
could be the potential loop-hole as QGP signal, as
it is very difficult to obtain its dramatically large yield
(compared to p+p) by pure binary final state hadronic
interactions without invoking additional pre-hadronic
mechanisms \cite{SB99,Va01}; see, however, also \cite{AA00,PKL01a}.

Cassing made a significant step forward very recently
in trying to implement such `back-reactions'
withinn a present transport code.
First results concerning the production of anti-protons
at AGS and SPS energies are quite impressive \cite{Ca01}.
There is also a clear hint at AGS energies
of enhanced anti-$\Lambda $ production:
E917 \cite{E917} had reported very recently a large value for the
$\bar{\Lambda }/\bar{p} $ ratio of $3.6^{+4.7}_{-1.8}$
for central Au+Au collisions, which confirms
earlier, but indirect measurements by E864 and E878, and by E859
for the lighter Si+Au collision at 14.6 AGeV.
Here it is mandatory to first understand the $\bar{p}$-production as
reported by e.g. E802 and to see whether the multi-pionic
channel is dominant. This seems to be indeed the case \cite{Ca01}.
The enhanced production of anti-$\Lambda $s compared to
anti-protons at AGS energies one can
understand in a way that one assumes that their annihilation cross section
on baryons is just slightly smaller \cite{CG01}. But this is speculation
at present. Unfortunately, there is no data for $\bar{\Xi }$ at AGS.
A clean and detailed measurement of all antihyperons
represents an excellent opportunity for future heavy ion facilities
at an energy upgraded GSI.

\section{Conclusions: `Where are we $\ldots $?'}
\label{sec:Concl}
I have tried to review three of the early great ideas of
why strange particles can probe interesting, collective
physics of strong interactions in a macroscopic system.
For the theoretical guide of the discussion I had tried to review
and to give insight in the present and very detailed knowledge
of hadronic transport approaches. They do serve
as an excellent benchmark to understand the complete
nonequilibrium evolution, to test and implement, if possible, various
ideas and, if necessary, to really unravel new exciting physics beyond
standard transport.

Do we see a strongly reduced effective energy for the antikaons,
and a slightly enhanced one for the kaons at SIS energies?
All groups do need the potentials in one or the other way.
None of the many transport approaches can describe the various
data without invoking some sort of in-medium modifications.
So there is a good indication. However, there still remains
some important cross sections to be adressed and compared in detail
among the transport competitors in order to converge
to a consistent picture.

Do we have learned the nuclear EoS from kaon measurement?
There is a moderate indication that indeed one can extract
a soft and momentum dependent EoS from comparison with data.
Still, because of the various ambiguities in the cross sections,
which have to be adressed, I think the claim might be premature.
In any case cross correlations with other potential good observables
on the nuclear EoS should be consistently adressed with each individual
model if to come for a final consensus.

Last not least the historically intriguing question whether
we do see the QGP from abundant
strange particle production and, in particular, from antihyperons?
As elaborated, I believe that the overall strangeness
discussed is produced at the very early nonequilibrium stage.
As this seems to be true for all energies this does not prove deconfinement.
How does then strange flavor redistribution to rare particles happen.
At SIS energies the rare antikaons are not produced directly but stem from
secondary flavor exchange reaction of a hyperon on a pion.
An alternative reasoning could be true for making the rare antihyperons
at SPS and AGS energies: Intriguing, novel, multi-mesonic collapses to form
baryon-antibaryon pairs could efficiently do the job.
Hence, antihyperons can be produced by more collective and complicated, yet
hadronic kinetic reactions.

\section*{Acknowledgements}
Part of the results presented had been done in various collaborations
with E.~Bratkovskaya, W.~Cassing, J.~Geiss, S.~Leupold and U.~Mosel.
I am in particular indebted to J.~Aichelin, E.~Bratkovskaya and
H.~Oeschler for the exhaustive discussions on the various
experimental and theoretical results and problems concerning
the production of strange particles within
the relativsitic heavy ion physics program with the SIS at GSI/Darmstadt.
Further discussions with S.~Bass, F.~Becattini,
M.~Bleicher, W.~Cassing, D.~R\"ohrich and C.~Sturm are also
acknowledged.
This work has been supported by BMBF and GSI Darmstadt.

\section*{References}

\end{document}